\documentclass{amsart}
\usepackage[latin1]{inputenc}

\usepackage{amssymb}
\usepackage{amsmath,enumerate,rotate}
\usepackage{amssymb,latexsym}
\usepackage{epsfig}
\usepackage{graphicx}
\usepackage[english]{babel}
\usepackage{natbib}

\makeatletter

\newtheorem{definition}{Definition}[section]

\newtheorem{lemma}[definition]{Lemma}
\newtheorem{proposition}[definition]{Proposition}
\newtheorem{corollary}{Corollary}
\newtheorem{remark}{Remark}
\newtheorem{example}{Example}

\newcommand\al{\alpha}

\def\CX {\mathcal{X}}
\def\CF { \mathcal{F}}
\def\CL { \mathcal{L}}
\def\CM { \mathcal{M}}

\def\CB {\mathcal{B}}
\def\CP { \mathcal{P}}

\def\CL { \mathcal{L}}

\def\CS {\mathcal{S}}

\def\CP {\mathcal{P}}

\def\CT {\mathcal{T}}

\def\a {\alpha}
\def\t {\tau}
\def\d {\delta}
\def\th {\theta}
\def\s {\sigma}

\def\eps {\epsilon}

\def \TH {\Theta}
\def\a {\alpha}

\def\t {\tau}

\def\lm {\lambda}

\def\J {\mathbb{I}}
\def\M {\mathbb{M}}

\def\N {\mathbb{N}}

\def\E {\mathbb{E}}

\def\RE {\mathbb{R}}

\def\N {\mathbb{N}}
\def\P {\mathbb{P}}

\def\X {\mathbb{X}}
\def\D {\mathbb{D}}
\def\A {\mathbb{A}}

\def \te {\tilde e}
\def \tp {\tilde p}

\def \tm {\tilde \mu}

\def \fine{\diamondsuit}

\def \va {\xi}

\def\be{\begin{equation}}
\def\ee{\end{equation}}

\newcommand\LC[2]{{\CL}_{ #1\mid #2}} 
\newcommand\LM[1]{{\CL}_{ #1}} 

\def \QnN  {\LC{\te_N}{\xi(n)} }
\def \QnNp   {\LC{\te_N}{\xi(n),\tp} } 
\def \Qxn     { \LC{\tp}{\xi(n)} }   

\def \zi { z_i^{(p)} }

\usepackage{verbatim}

\begin{document}

\title[Comparing posterior distributions]{ Quantitative comparisons between finitary posterior distributions
and Bayesian posterior distributions}

\thanks{
\\
Research partially supported by Ministero dell'Istruzione, dell'Universit\`a e della Ricerca
 (MIUR grant 2006/134526) the IMATI (CNR - Pavia, Italy).\\
}

\author{
Federico Bassetti 
 }

\address{Universit\`a degli Studi di  Pavia, Dipartimento di Matematica, via Ferrata 1, 27100 Pavia, Italy}

\email{federico.bassetti@unipv.it} 


\begin{abstract}
{
The main object of Bayesian statistical inference is the 
determination of posterior distributions. 
Sometimes these laws are given for quantities devoid of empirical value. 
This serious drawback vanishes when one confines oneself to 
considering a finite horizon framework.  
However, assuming infinite exchangeability  
gives rise to fairly tractable {\it a posteriori} 
quantities, which is very attractive in applications. 
Hence, with a view to a reconciliation between these two aspects
of the Bayesian way of reasoning, in this paper
 we provide quantitative comparisons between posterior distributions of 
finitary parameters and posterior distributions of allied parameters appearing in 
usual statistical models. 
}

\end{abstract}


\maketitle


\section{Introduction}\label{intro}

In the Bayesian reasoning the  assumption of infinite exchangeability 
gives rise to fairly tractable {\it a posteriori} 
quantities, which is  very attractive in real applications. 
If observations  form an infinite exchangeable sequence 
of random variables, de Finetti's representation 
theorem states that they are conditionally independent and identically distributed,
given some random parameter, and  the distribution of this random parameter is the 
center of the current Bayesian statistical inference. 
The theoretical deficiency of this approach lies  
in interpreting these parameters. 
In  fact, as pointed out for the first time by de Finetti (see \cite{deFinetti30},\cite{deFinetti37} 
and also \cite{Bassettiregazzini2008}), 
parameters ought to be of such a
nature that one should be able to
acknowledge at least the theoretical possibility of experimentally
verifying whether hypotheses on these parameters are true or false.
A closer look to the usual Bayesian procedures shows that
Bayesian statisticians often draw inferences (from
observations) both to empirical (i.e. verifiable) and to non
empirical hypotheses. 
To better understand this point, it is worth stating a more complete formulation of the 
already mentioned de Finetti's
representation theorem: {\it  A  sequence $(\va_n)_{n \geq 1}$ of
 random elements taking values in some suitable measurable space $(X,\CX)$
(e.g. a Polish space), is exchangeable if and only if  the {\it empirical distribution}
\[
\te_n(\cdot)=\frac{1}{n} \sum_{i=1}^n \d_{\va_i}(\cdot)
\]
converges in distribution to a random probability $\tilde p$ with
probability one and  the $\va_n$s turn out
to be conditionally independent given $\tilde p$, with common 
distribution $\tilde p$. }
Hence, it is $\tilde p$ 
that takes  the traditional
role of parameter in Bayesian modeling. However, since $\tilde p$ is a limiting entity of mathematical nature, hypotheses 
related to it might be devoid of empirical value. 
It is clear that this drawback vanishes when one confines oneself to considering a finite horizon framework,
in which $\te_N$, that is always (at least ideally) observable, takes the place of
$\tilde p$. In this way one preserves the hypothesis of exchangeability, which is  quite natural 
in many statistical problems, but one avoids the problem of assessing probability law to unobservable entities.  
In particular, in this context,  the conditional distribution of the
empirical measure $\tilde e_N$ given $\xi(n):=(\xi_1, \dots,
\xi_n)$ ($n < N$)  takes the place of the conditional distribution
of $\tilde p$ given given $\xi(n)$, i.e. the usual posterior distribution 
of the Bayesian (nonparametric) inference. 

Even if, in view of de Finetti's representation, 
the parameter corresponding to the so--called ``unknown distribution''
(i.e. $\tilde p$) is the limit, as $N \to +\infty$, of empirical distribution, 
it should be emphasized that in the Bayesian practice two conflicting aspects sometimes 
occur. On the one hand, statistical inference ought to concern finitary and, 
therefore, observable entities whereas, on the other hand, simplifications of a technical nature
can generally be obtained by dealing with (parameters defined as function
of) the ``unknown distribution'' $\tilde p$. Hence,  
it is interesting  to compare the conditional distribution of $\te_N$ given $\xi(n)$ with the conditional 
distribution of $\tp$ given $\xi(n)$, when $(\xi_k)_{k\ge 1}$ is an infinite exchangeable sequence directed by $\tilde p$.
This is the aim of this paper,
that can be thought of as a continuation of the papers \cite{BisBas1} and \cite{pippo}, where
specific forms of (finitary) exchangeable laws have been defined and studied in terms of 
finitary statistical procedures.

The rest of the paper is organized as follows. Section
\ref{secFinitaryProc} contains a brief overview of the finitary
approach to statistical inference together with some examples. 
Sections   \ref{Merging1} and  \ref{Merging2} deal
with the problem of quantifying the discrepancy between the conditional 
law of $\tilde e_N$ given $\xi(n)$ and
the conditional law of $\tilde p$ given $\xi(n)$ .

To conclude these introductory remarks it is worth mentioning  \cite{DiaconisFreedman80}, 
which, to some extent,  is connected with our present work. 
In point of fact, in \cite{DiaconisFreedman80}, Diaconis and Freedman provide 
an optimal bound for the total variation distance between the law of $(\xi_1,\dotsc,\xi_n)$ and the 
law of $(\zeta_1,\dotsc,\zeta_n)$,  $(\xi_1,\dotsc,\xi_N)$ being a given finite exchangeable sequence and $(\zeta_k)_{k\ge 1}$ a suitable infinite exchangeable sequence.

\section{Finitary statistical procedures}\label{secFinitaryProc}
As said before, we
 assume that the process of observation can be represented as   an infinite
exchangeable sequence  $(\xi_k)_{k \geq 1}$  of random elements defined on a probability space
$(\Omega,\CF,P)$ and  taking values in a  complete separable metric
space $(X,d)$, endowed with its Borel $\s$--field $\CX$.
Let $\P_0$ be a subset of the set $\P$ of all probability
measures on $(X,\CX)$ and let $t:\P_0 \to \TH$ be a parameter of interest, 
$\TH$ being a suitable parameter space endowed with a $\s$--field. 
From a finitary point of view, a statistician must 
focus his attention on empirical versions
\(
t(\te_N)
\)
of the more common  parameter $t(\tilde p)$. 

It might be useful, at this stage, to recast the decision theoretic formulation of a statistical problem 
in finitary terms. Usually one assumes that the statistician has a set $\D$ of {\it decision rules} 
at his disposal and that these rules are defined, for any $n \leq N$, as functions from $\X^n$ to some set $\A$ of actions. Then one considers a {\it loss function} $L$, i.e.
a positive real--valued function on $\TH \times \A$, such that $L(\th,a)$ represents the loss
when the value of $t(\te_N)$ is $\th$ and the statistician chooses action $a$.
It is supposed that
\[
r_N(\delta(\va(n)) ):=\E[ L(t(\tilde e_N), \delta(\va(n)))|\va(n)]
\]
is finite for any $\delta$ in $\D$ 
and, then,  $r_N(\cdot)$
is said to be the {\it a posteriori Bayes risk} of $ \delta(\va(n))$. Moreover, a {\it Bayes rule}
is defined to be any element $\delta_{FB}$ of $\D$ such that
\[
r_N(\delta_{FB}(\va(n)) )=\min_{\delta \in \D} r_N(\delta(\va(n)) )
\]
for any realization  of $\va(n)$.
We shall call such a Bayes rule  {\it finitary Bayes estimator} in order to distinguish  it from the more common Bayes estimator
obtained by minimizing
\[
r(\delta(\va(n)) ):=\E[ L(t(\tilde p), \delta(\va(n)))|\va(n)].
\]
While the law of the latter estimator is determined by the posterior distribution, that is the conditional distribution of $\tp$ given $\xi(n)$, 
the law of a finitary Bayes estimator is determined by the "finitary" posterior distribution,
that is the conditional distribution of $t(\te_N)$ given $\xi(n)$.
A few simple examples will hopefully clarify the connection between the finitary Bayesian procedures and the  usual
Bayesian ones.
In all the  examples we shall present, observations are assumed to be real--valued, that is $(X,\CX)=(\RE,\CB(\RE))$, 
the space of actions is some subset of $\RE$  and the loss function is quadratic, i.e. 
$L(x,y)=|x-y|^2$. It is clear that, under these hypotheses,  
\[
\delta_{FB}(\va(n))= \E[t(\tilde e_N)|\xi(n)]
\]
and the usual Bayes estimator is given by
\[
\E[t(\tp)|\xi(n)].
\]

\begin{example}[Estimation of the mean]
Suppose the statistician has to estimate the mean under the squared
error loss, i.e. the functional of interest is $t(p):=\int_\RE x p(dx)$.
The usual Bayes estimator is
$$ 
\hat \mu_{n}:=\E[\xi_{n+1}|\xi(n)] 
$$
while the "finitary
Bayes" estimator is
\[
\hat \mu_{FB}=\frac{n}{N} \bar \mu_n+\frac{N-n}{N}\hat \mu_{n},
\]
where
$$
\bar \mu_n= \frac{1}{n} \sum_{i=1}^n\ \xi_i.
$$
Note that in this case the finitary Bayes estimator is a convex combination of the usual Bayes estimator
with the empirical (plug--in) estimator $\bar \mu_n$.
\end{example}

\begin{example}[Estimation of the variance]
Now, consider the estimation of the variance  $t(p)=\int_\RE x^2 p(dx) -(\int_\RE x p(dx))^2$, 
 under
the squared error loss. In this case the space of actions is
$\RE^+$ and the usual Bayes estimator is
\[
\hat \s^2_{n}:= \hat s_n^2 -  \hat c_{1,2,n}
\]
where
\[
\hat s_n^2 :=E[\xi_{n+1}^2|\xi(n)] \qquad \text{and} \qquad \hat c_{1,2,n}:=E[\xi_{n+1} \xi_{n+2}|\xi(n)].
\] 
Some computations show that the "finitary Bayes" estimator is
\[
\hat \s^2_{FB}=\frac{n}{N} \bar s_n^2+\frac{N-n +n/N-1}{N} \hat s_n^2
-\frac{n^2}{N^2} \bar c_{1,2,n} - \frac{(N-n)(N_n-1)}{N^2} \hat c_{1,2,n}
-\frac{2(N-n)n}{N^2} \bar \mu_n \hat \mu_n
\]
where
\[
\bar s_n^2 :=\frac{1}{n} \sum_{i=1}^n \xi_i^2 \qquad \text{and} \qquad \bar c_{1,2,n}:= \frac{1}{n^2} \sum_{i=1}^n\sum_{j=1}^n 
\xi_i\xi_j.
\]
\end{example}

\begin{example}[Estimation of the distribution function]
Assume one has to estimate $t(p)=F_p(y)=p\{(-\infty,y] \}$, 
where $y$ is a fixed real number. 
Under the square loss function, the classical Bayes estimator is
\[
\E(\J_{(-\infty,y]}(\xi_{n+1})|\xi(n))
\]
while the "finitary Bayes" estimator is
\[
\hat F_{FB}(y)=\frac{n}{N}
E_n(y)+\frac{N-n}{N} \E(\J_{(-\infty,y]}(\xi_{n+1})|\xi(n))
\]
where $E_n(y)=\frac{1}{n} \sum_{i=1}^n\J_{(-\infty,y]}(\xi_i)$.
\end{example}

\begin{example}[Estimation of the mean difference]
Estimate the Gini  mean difference 
\[
t(p)=\Delta(p)=\int_{\RE^2} |x-y|p(dx)p(dy)
\]
under the squared error loss. 
The usual Bayes estimator is
\[
\E (|\xi_{n+1}-\xi_{n+2}| |\xi(n) )
\]
while the "finitary Bayes" estimator is
\[
\begin{split}
\E(\Delta(\tilde e_N)|\xi(n))&=\frac{n^2}{N^2} \Delta_n +
\frac{(N-n)^2-(N-n)}{N^2} \E (|\xi_{n+1}-\xi_{n+2}| |\xi(n) )
\\&+\frac{2(N-n)}{N^2} \sum_{j < n}\E(|\xi_j-\xi_{n+1}| |\xi(n)), \\
\end{split}
\]
where
\[
\Delta_n:=\frac{1}{n^2} \sum_{i,j  \leq n}|\xi_i-\xi_j|.
\]
\end{example}

It is worth noticing that in all the previous examples when $N$ goes to $+\infty$ the 
"finitary Bayes" estimator converges to the usual Bayes estimator, while the 
"finitary Bayes" estimator becomes the usual plug--in frequentistic estimator if $n=N$ .

\section{Comparison between posterior distributions of means}\label{Merging1}
Let $Q$ be the probability distribution of $\tp$. 
Then  $Q$ turns out to be a probability measure on $\P(X)$.
Without loss of generality consider $\P=\P(X)$ endowed with
a bounded metric $\lambda$ which induces the weak convergence on
$\P$ (e.g. the Prohorov metric), and set $\CP$ for its
Borel $\s$--field. 
 In what
follows, if necessary, expand $(\Omega,\CF,P)$ in order to contain
all the random variables needed and, for any random variable $V$,
let $\LM{V}$  designate the  probability distribution of $V$ and,
for any other random element $U$, by $\LC{V}{U}$  denote some conditional probability 
distribution for $V$ given $U$.
In particular, $\QnN$ will denote (a version of) the
 conditional distribution of $\te_N$ given
$\xi(n):=(\xi_1,\dots,\xi_n)$ and $\Qxn$  will stand for (a version of) the
 conditional distribution of $\tilde p$ given $\xi(n)$, i.e.
the so--called posterior distribution of $\tilde p$. Such
distributions exist since $(\P,\lambda)$ is Polish.

As already said, the main goal of this paper is comparing $\QnN$ with $\Qxn$. 
We start by comparing posterior means. Indeed, as we have seen in the previous section,
  the posterior
mean of a function $f$ appears in many natural statistical estimation
problems. 
For the sake of notational simplicity, for  any  measurable real--valued function $f$, 
set
\[
\te_N(f):=\int_X f(x) \te_N (dx)=\frac{1}{N} \sum_{i=1}^N f(\xi_i)
\]
and
\[
\tp \,(f)\,:=\int_X f(x)\, \tp \,(dx).
\]

First of all we prove this very simple 

\begin{proposition}\label{P5.1} Given a real--valued measurable function
$f$, if \( P\{ \tp(|f|) < +\infty  \}=1 \), then $\te_N(f)$ converges
in law to $\tp(f)$ {\rm(}as $N \to +\infty${\rm)}. Analogously,
 $ \LC{\te_N (f)}{\xi(n)}$ converges weakly
{\rm(}almost surely{\rm)} to   $\LC{\tp (f)}{\xi(n)}$.
\end{proposition}

{\it Proof.} Let  $\phi$ be a bounded continuous function with
$c=\|\phi\|_\infty$. Then $\phi(\te_N) \leq c$. Now,
$\E(\phi(\te_N(f)|\tp)$ converges almost surely to
$\E(\phi(\tp(f))|\tp)$. To see this, note that, conditionally on  $\tp$, $\te_N(f)$
is a sum of independent random variables with mean $\tp(f)$ and absolute moment $\tp(|f|)$, and, since 
 $\tp(|f|)$ is almost surely finite, the conditional law of $\te_N(f)$ given $\tp$
converges almost surely to $\tp(f)$, and hence also in law. Since
$|\E(\phi(\te_N(f)|\tp)| \leq c$ almost surely, to conclude the proof
it is enough to apply the dominated convergence theorem.
The second part of the theorem can be proved in the same way
conditioning with respect to
 $(\tp,\xi(n))$. $\quad \fine$

In order to give a  quantitative version of the previous
statement we resort to the so--called
Gini--Kantorovich--Wasserstein distance. Let $\P_1=\P_1(\RE^d)$ be the subset
of the set $\P=\P(\RE^d)$ of all probability measures on $\CB(\RE^d)$ defined by
$\P_1:=\{ p \in \P : \int_{\RE^d}\| x \|p(dx)<+\infty \}$, where $\|\cdot\|$ denotes the Euclidean norm on
$\RE^d$. For every couple of probability measures $(p,q)$ in
$\P_1 \times \P_1$ the Gini-Kantorovich-Wasserstein distance (of order one) between
$p$ and $q$ is defined by
\[
w_1(p,q):=\inf \left\{ \int_{\RE^{2d}} \|x-y \| \gamma(dx dy) :
\gamma \in \CM(p,q) \right\}, 
\]
$\CM(p,q)$ being the class of all probability measures on
$(\RE^{2\times d},\CB(\RE^{2\times d}))$ with marginal distributions $p$ and $q$.
For a general definition of the Gini-Kantorovich-Wasserstein distance and its properties see, e.g., 
\cite{Rachev1991}.
If $Z_1$ and $Z_2$ are
two random variables with law $p$ and $q$ respectively,
$w_1(Z_1,Z_2)$ will stand for $w_1(p,q)$.

\begin{proposition} Given a real--valued measurable function
$f$ such that $\E[f(\xi_1)^2]<\infty$, then
\[
w_1(\te_N(f),\tp(f)) \leq \frac{1}{\sqrt{N}} \left ( \E|f(\xi_1)-\,
\tp \,(f) \,|^2 \right )^{1/2} \leq \frac{2}{\sqrt{N}}\sqrt{\E[f(\xi_1)^2]}.
\]
Moreover,
\[
\begin{split}
w_1(\LC{\te_N (f)}{\xi(n)},\LC{\tp (f)}{\xi(n)}) &\leq \frac{n}{N}
\left( \frac{1}{n}\sum_{i=1}^n f(\xi_i) \,+\E[ \,\tp \,
(f)|\,\xi(n)] \right )
\\ & +\frac{2}{\sqrt{N-n}} \left ( \E[f(\xi_{n+1})^2|\xi(n)]
\right )^{1/2}
 \qquad (a.e.). \\
\end{split}
\]
\end{proposition}

{\it Proof.} Applying a well--known conditioning argument,  note that
\[
\begin{split}
w_1(\te_N(f),\tp(f)) &\leq \E|\te_N(f)-\tp(f)| \\
&= \E[ \E (  |\te_N(f)-\tp(f)| |\tp)] = \E\Big [\E\Big (
\Big |\frac{1}{N}\sum_{i=1}^N  f(\xi_i) -\int f\tp(dx)\Big | \Big |\tp\Big)\Big]
\\ & \qquad \text{(by the  Cauchy-Schwartz  inequality)} \\&
\leq \frac{1}{\sqrt{N}} \E\left[\E\Big[\Big(f(\xi_1)-\int f \tp\Big)^2
\Big|\tp\Big]^{1/2} \right] \\
& \qquad \text{(by the Jensen  inequality)} \\& \leq \frac{1}{\sqrt{N}}
\E\left[(f(\xi_1)-\int f \tp)^2
 \right]^{1/2}. \\
\end{split}
\]
Clearly, $\E[(f(\xi_1)-\int f \tp)^2]^{1/2} \leq (2 (\E[f(\xi_1)^2+ (\int f \tp)^2  )^{1/2}$
and, by the Jensen inequality, $\E(\int f \tp)^2 \leq  \E(\int f^2 \tp)=\E[f(\xi_{1})^2]$.
As for the second part of the proposition, first note that
\[\begin{split}
w_1(\LC{\tp (f)}{\xi(n)},\LC{\te_N (f)}{\xi(n)})
&\leq \frac{N-n}{N} \E \left[
\frac{1}{N-n} \sum_{i=n+1}^N f(\xi_i)-\tp \, (f)\,   \Big  |\xi(n) \right]
\\&+\frac{n}{N} \E  \left [ \frac{1}{n} \sum_{i=1}^n f(\xi_i)-\tp \, (f)\,
\Big |\xi(n)  \right].\\
\end{split}
\]
Now, take the conditional expectation given $(\tp, \xi(n)) $ and
use again the Cauchy-Schwartz  inequality to obtain
\[
\begin{split}
\E   [ \frac{1}{N-n} &\sum_{i=n+1}^N f(\xi_i)-\tp \, (f)\,   |\xi(n) ]
\\ &\leq \frac{1}{\sqrt{N-n}} \E  [ \E\big ( | f(\xi_{n+1})-\tp \,(f)|^2\,
|\tp,\xi(n) \big)\big |\xi(n)  ]^{1/2}. \\
\end{split}\]
Finally, to complete the proof, apply  the Jensen inequality and argue as in the previous part of the proof. $\quad \fine$

Of course, the mean is not the unique interesting functional which
appears in statistical problems. For instance,
 statisticians  frequently deal with functionals of the form
\[
t_1(p)=\int_{X^k} f(x_1,\dots,x_k) p(dx_1)\dots p(dx_k),
\]
or even of the form
\[
 t_2(p)= \text{argmin}_{\th \in \TH} \int_{X^k} f_\th(x_1,\dots,x_k) p(dx_1)\dots
p(dx_k).
\]
Think, for example, of the variance or of the median of a probability measure, respectively.

It is immediate to generalize Proposition \ref{P5.1} according to 
\begin{proposition}\label{P5.5} Given a  measurable function
$f:X^k \to \RE$ such that  
\[P\{ \int_{X^k} |f(x_1,\dots,x_k)|
\tp(dx_1)\dots \tp(dx_k) < +\infty \}=1, 
\]
then $ t_1(\te_N)$
converges in law to $t_1(\tp)$ {\rm(}as $N \to +\infty${\rm)}.
Analogously,
 $\LC{\xi(n)}{t_1(\te_N)} $  converges weakly  
{\rm(}almost surely{\rm)} to   $\LC{\xi(n)}{t_1(\tp)}$.
\end{proposition}

As far as functionals of the type of $t_2$ are concerned, the
situation is less simple. As a general strategy, one could
apply the usual {\it argmax argument}. See, e.g., \cite{vanderVaartWellner1996}. 
To do this,  set
\[
\M_N(\th):= \int_{X^k} f_\th(x_1,\dots,x_k) \te_N(dx_1)\dots
\te_N(dx_k),
\] 
\[\M(\th):=\int_{X^k} f_\th(x_1,\dots,x_k) \tp(dx_1)\dots
\tp(dx_k)\] 
and $\th_N=t_2(\te_N)$. 
Assume that $\TH$ is a subset of
$\RE^d$ and, for every 
$T \subset \RE^d$ define
the set $l^\infty(T)$ of all measurable functions $f:T \to \RE$ satisfying
\[
\|f\|_T:=\sup_{t \in T}|f(t)|< +\infty.
\]
A version of the argmax theorem (Theorem 3.2.2 in  \cite{vanderVaartWellner1996}) 
implies that: {\it If 
$\M_N$ converges in
law to $\M$ in $l^\infty(K)$ for every compact set $K \subset
\RE^d$, if almost all sample paths $\th \mapsto \M(\th)$ are lower
semi-continuous and possess a unique minimum at a random point $\hat
\th=t_2(\tp)$, and if $( \th_N)_{ N \geq 1}$ is tight, then  $\th_N$ converges in
law to $\hat \th$.}

As for the first hypothesis, that is $\M_N$ converges in
law to $\M$ in $l^\infty(K)$ for every compact set $K \subset
\RE^d$,
one can resort to Theorems 1.5.4 and 1.5.6 in
\cite{vanderVaartWellner1996}. Such theorems  
imply   that if $(\M_N(\th_1),\dots,\M_N(\th_k))$ converges in law to
$(\M(\th_1),\dots,\M(\th_k))$ for every $k$ and every $(\th_1,\dots,\th_k)$
in $K^{k}$ and if, for every \( \epsilon\) and \( \eta > 0\), there is
a finite partition \( \{T_1,\dots, T_N \} \) of $K$ such that
\begin{equation}\label{tightwellner}
\limsup_N P \{\sup_i \sup_{h_1,h_2 \in
T_i}| \M_N(h_1)-\M_N(h_2)|>\epsilon \} < \eta
\end{equation}
then $\M_N$ converges in
law to $\M$ in $l^\infty(K)$ for every compact set $K \subset
\RE^d$.
Hence, 
one can try to show that 
\[
|f_{\th_1}(x_1,\dots,x_k)-f_{\th_2}(x_1,\dots,x_k)| \leq
g(\|\th_1-\th_2 \|_2) \phi(x_1,\dots,x_k)
\]
for some continuous function  $g$, with $g(0)=0$, and some function $\phi$ such
that for some $\th_0$
\[
P\left \{ \int_{X^k} [\phi(x_1,\dots,x_k)
+|f_{\th_0}(x_1,\dots,x_k)|] \tp(dx_1)\dots \tp(dx_k) < +\infty
\right \}=1.
\]
If these conditions hold, then the convergence of $\M_N$ to $\M$ is easily proved, 
whereas both tightness of $(\th_N)_{ N \geq 1}$
and uniqueness of $\hat \th$ require additional assumptions.

Here is an example, where $med(p)$ denotes the median of the distribution $p$.

\begin{proposition}
Let $M_N=med(\te_{2N+1)})$,  that is $M_N=\xi_{(N+1)}$ if $\xi_{(1)}
\leq \dots \leq \xi_{(2N+1)}$. If
\[
P\left \{ \int_\RE |x| \tp (dx) < +\infty \right \}=1 \qquad \text{and} \qquad
 P\{ med(\tp) \, \text{is unique} \, \}=1,
\]
then $M_N$ converges in law to $med(\tp)$ as $N$ diverges. Analogously,
if, for some $n<N$,
\[
\xi(n) \mapsto P\left \{ \int_\RE |x| \tp (dx)<+\infty \,\, \Big | \,\,\xi(n) \right \}=1 \qquad (a.e.)
\]
and
\[
\xi(n) \mapsto  P\{ med(\tp) \, \text{is unique} \,|\xi(n)  \}=1, \qquad (a.e.),
\]
then $\LC{M_N}{\xi(n)}$ converges weakly (almost surely) 
to $\LC{med(\tp)}{\xi(n)}$ as $N$ diverges.
\end{proposition}

{\it Proof.} 
In this case
\[
\M_N(\th)= \int_\RE |x-\th| d \te_{2N+1}
\]
and 
\[
\M(h)=\int_\RE |x-\th| d \tp.
\]
Since $P\left \{ \int_\RE |x| \tp (dx) <+\infty \right \}=1$, 
from Proposition \ref{P5.5} we get that 
$(\M_N(\th_1),\dots,\M_N(\th_k))$ converges in law to
$(\M(\th_1),\dots,\M(\th_k))$ for every $k$ and every $(\th_1,\dots,\th_k)$.
Moreover
\[
|\M_N(\th_1)-\M_N(\th_2) | \leq |\th_1 -\th_2|,
\]
hence (\ref{tightwellner}) is verified. 
It remains to prove the tightness of $(M_N)_{N \geq 1}$.
First of all observe that if $X_1,\dots,X_{2N+1}$ are i.i.d random variables with 
common distribution function $F$ then the distribution function of the median of 
$X_1,\dots,X_{2N+1}$ is given by
\[
x \mapsto \sum_{k=N+1}^{2N+1} {2n+1 \choose k} F^{k}(x)(1-F(x))^{2N+1-k}
=\frac{1}{B(N+1,N+1)} \int_0^{F(x)}t^N(1-t)^{N}dt,
\]
where $B$ is the Euler integral of the first kind (the so--called beta function).  
Hence, denoting by $\tilde F(x)$ the distribution function of $\tilde p$
 and setting
 \[
 H_x(\t)=P\{ \tilde F(x) \leq \tau\},
 \]
 it follows that
 \[
 \begin{split}
 P\{M_N \leq x  \}&= \E\left ( B(N+1,N+1)^{-1} \int_0^{\tilde F(x)}t^N(1-t)^{N}dt \right) \\
 &= B(N+1,N+1)^{-1} \int_0^1  \int_0^{\tau}t^N(1-t)^{N}dt dH_x(\t)\\
 &= B(N+1,N+1)^{-1} \int_0^1 t^N(1-t)^{N} [1-H_x(t)]dt .\\
  \end{split}
  \]
Now, by the Markov inequality,  
\[
 [1-H_x(t)] = P\{ \tilde F(x) >t  \} \leq \frac{1}{t} \E[\tilde F(x) ]=P\{ \xi_1 \leq x  \},
\]
hence, 
\[
  P\{M_N \leq x  \} \leq P\{ \xi_1 \leq x  \}  B(N+1,N+1)^{-1} \int_0^1 t^N(1-t)^{N} dt 
=P\{ \xi_1 \leq x  \} \frac{2N+1}{N}.
\]
In the same way, it is easy to see that
\[
\begin{split}
P\{M_N < x  \}& =1-
P\{M_N \leq x  \}= 1-B(N+1,N+1)^{-1} \int_0^1 t^N(1-t)^{N} [1-H_x(t)]dt \\
&=B(N+1,N+1)^{-1} \int_0^1 t^N(1-t)^{N} H_x(t)dt \\
&=B(N+1,N+1)^{-1} \int_0^1 t^N(1-t)^{N} H_x(1-t)dt
\end{split}
\]
and hence
\[
P\{M_N > x  \} \leq \frac{2N+1}{N} P\{\xi_1 \geq x    \}.
\]
With these inequalities it is immediate to prove the tightness of 
$(M_N)_{N \geq 1}$.  
The proof of the  second part of the proposition is analogous. 
$\quad \fine$

\section{Comparing posterior distributions of random probabilities}\label{Merging2}

We now turn our attention to the comparison of $\QnN$ with $\Qxn$. 
We shall use the Gini-Kantorovich-Wasserstein distance on the space of all probability measures. 
The Gini-Kantorovich-Wasserstein 
distance of order $1$ (relative to a metric $\lm$)
between two probability
measures, say $(Q_1,Q_2)$, defined on $(\P,\CP)$  
 is 
\[
W_1(Q_1,Q_2):= \inf \left \{ \int_{\P^2} \lambda(p_1,p_2) \Gamma(dp_1
dp_2):  \Gamma \in M(Q_1,Q_2) \right \}
\]
where $M(Q_1,Q_2)$  is the set of all probability measures on
$(\P\times\P,\CP\otimes \CP)$ with marginals $Q_1$ and $Q_2$.
Here, it is worth recalling that $W_1$ admits the following dual representation
\begin{equation}\label{dualrep}
\begin{split}
W_1(Q_1,Q_2)= \sup & \Big \{ \int_{\P} f(p)(Q_1(dp)-Q_2(dp)) ;
\\ &f:\P \to \RE,\,\,\,\, |f(p)-f(q)| \leq \lambda(p,q) \,\, \,\,
\forall \,\, p,q \,\,\in \,\, \P
\Big \}. \\
\end{split}
\end{equation}
See, e.g., Theorem 11.8.2 in \cite{Dudley2002}.
The main
goal of this section is to give explicit upper bounds for the random variable
$W_1(\QnN,\Qxn)$.

\subsection{A first bound for the posterior distributions}
There is a large body of
literature on the rate of convergence to zero (when $N$ diverges) of
\[
E_N(p):=\E \left[ \lambda \left (p,\nu_{N}^{(p)} \right)\right],
\]
where
$\nu_{N}^{(p)}:=\sum_{i=1}^N \delta_{z^{(p)}_i}/N$ and  
$(z^{(p)}_i)_{i \geq 1}$
is a sequence of independent and identically distributed (i.i.d.)
random variables taking values in $X$, with common probability measure
$p$. See, for instance,  \cite{Alexander1984},  \cite{Dudley1968} and \cite{Kalashnikov1990}.
 The next lemma shows how these well--known results can be used 
to get a bound for $W_1(\Qxn,\QnN)$.

\begin{lemma}\label{L0} Assume
that $\lambda$ is bounded and satisfies 
\begin{equation}\label{qlin}
\lambda(p,\eps p_1 +(1-\eps)p_2) \leq \eps \lambda(p,p_1)
+(1-\eps)\lambda(p,p_2)
\end{equation}
for every $\eps$ in $(0,1)$
and every $p,p_1,p_2$ in $\P$.
Moreover, let $K:=\sup \{\lambda(p,q) :(p,q) \in \P^2\}$. Then
\begin{equation}\label{1bound}
W_1(\Qxn,\QnN) \leq \int_\P E_{N-n}(p) \Qxn(dp)+ \frac{n K}{N}
\end{equation}
holds true for $P$--almost every $\xi(n)$.
\end{lemma}

{\it Proof.} First of all, note that for every $A$ in $\CP$
\[
\QnN(A)= \int_\P \QnNp(A) \Qxn(dp)
\]
where , according to our notation, $\QnNp$ denotes (a version of) the  conditional
distribution of $\te_N$ given $(\xi(n),\tp)$. Hence, from the dual
representation (\ref{dualrep}) of $W_1$ it is easy to see that
\[
W_1(\Qxn,\QnN) \leq \int_\P W_1 \left(\delta_p, \QnNp\right)
\Qxn(dp).
\]
Now, write
\[
\te_N=\frac{n}{N}\te_{n} + \frac{N-n}{N} \te_{N,n}
\]
with $\te_{N,n}=\sum_{i=n+1}^N \delta_{\xi_i}/(N-n)$, and observe
that $\te_{n}$ and $\te_{N,n}$  are conditionally independent given
$\tilde p$. Moreover, $\te_{N,n}$ has the same law of $\te_{N-n}$
and $W_1 \left(\delta_p, Q\right)=\int_{\P} \lambda(p, q) Q(dq)$.
Hence,
\[
\begin{split}
W_1 \left(\delta_p, \QnNp\right)&=
 \int_{\P} \lambda(p, q) \QnNp(dq) \\
 &= \E\left [ \lambda\Big(p, \frac{1}{N}\sum_{i=1}^{N-n} \delta_{\zi}
+\frac{1}{N} \sum_{i=1}^n\delta_{\xi_i}
  \Big)  \right] \\
&\leq\frac{N-n}{N}
 \E\left [ \lambda\Big(p, \frac{1}{N-n}\sum_{i=1}^{N-n} \delta_{\zi}
  \Big)  \right]+ \frac{nK}{N}= \frac{N-n}{N}E_{N-n}(p)+\frac{nK}{N} . \qquad \qquad
  \end{split}
\]
The thesis follows from integration over $\P$ with respect to $\Qxn$.
$\fine$

In the next three subsections we shall use the previous lemma with
different choices of $X$ and $\lm$.

\subsection{The finite case} We start from the simple  case in which 
$X=\{a_1,\dots,a_k \}$. Here $\P$ can
be seen as the simplex
\[
\CS_k=\{ x \in \RE^k : 0 \leq x_i \leq 1, i=1\dots,k, \sum_{i=1}^k x_i =
1 \}.
\]
Define $\lambda$ to be the total variation distance,
i.e. $\lm(p,q)=\frac{1}{2}\sum_{i=1}^k |p(a_i)-q(a_i) |$. In point of fact, it should be noted that, 
since $X$ is finite, there is no difference between the strong and
the weak topology on $\P$. In this case,  for every $j=1, \dots, k$,
one has
\[
\te_N(a_j)= \sharp \{ i: \xi_i=a_j; \,\, 1 \leq i \leq N  \}/N.
\]
Now, denoting by $Z_i$  a binomial random variable of parameters
$(N-n,p_i)$ ($p_i:=p(a_i)$), we get
\begin{eqnarray*}
 \E \left[ \lambda\left(p,\nu_{N-n}^{(p)} \right)\right] &=&
\frac{1}{2(N-n)}\sum_{i=1}^k \E[|Z_i-(N-n)p_i|]\\
&\leq& \frac{1}{2(N-n)}\sum_{i=1}^k \sqrt{\E[|Z_i-(N-n)p_i|^2]}\\
&=&\frac{1}{2(N-n)}\sum_{i=1}^k \sqrt{(N-n)p_i(1-p_i)}=\frac{1}{2\sqrt{N-n}}\sum_{i=1}^k \sqrt{p_i(1-p_i)}\\
&\leq&\frac{k}{4\sqrt{(N-n)}}.
\end{eqnarray*}
\noindent Observing that $K:=\sup \{TV(p,q) :p,q \in\CS_k\}\leq 1$
and  that the total variation distance satisfies (\ref{qlin}),
Lemma \ref{L0} gives
\begin{proposition} If $X=\{a_1,\dots,a_k \}$, then
$$W_1(\Qxn,\QnN) \leq  \frac{k}{4\sqrt{N-n}}+\frac{n}{N}. $$
\end{proposition}

\subsection{The case $X=\RE$}
Passing to a general Euclidean space we first need to
choose a suitable metric $\lambda$. We recall that
if $p$ and $q$ belongs to $\P(\RE^d)$, the so--called 
bounded Lipschitz distance
(denoted by $\beta$) between $p$ and $q$
is
defined by
\[
\beta(p,q)=\sup \left \{\int_{\RE^d} f(x)[p(dx)-q(dx)]; 
\,\, f: \RE^d \to \RE, \,\, \|f\|_{BL}
\leq 1 \right \}
\]
where $\|f\|_{BL}:= \sup_{x \in \RE^d} |f(x)| + \inf_{(x,y) \in
\RE^d\times \RE^d} |f(x)-f(y)|/\|x-y\|$. See Section 11.3 in
\cite{Dudley2002}. Note that $\sup_{(p,q) \in \P} \beta(p,q) \leq 2$
and that $\beta$
satisfies 
$\beta(p,\eps p_1 +(1-\eps)p_2) \leq \eps \beta(p,p_1) +
(1-\eps)\beta(p,p_2)$ for every $\eps$ in $(0,1)$ and every $p$,
$p_1$, $p_2$ in $\P$. Recall also that $\beta$ metrizes the weak
topology (see, e.g., Theorem 11.3.3 in \cite{Dudley2002} ).
In what follows we take $X=\RE$ and $\lambda=\beta$. 
As a consequence of Lemma \ref{L0},
we get the next proposition in which, for every $p$ in $\P$, 
we set $F_p(x)=p\{(-\infty,x]\}$.

\begin{proposition}\label{P2.2} Let $X=\RE$ and $\lambda=\beta$. Set
$\Delta(p):=\int_\RE \sqrt{F_p(t)(1-F_p(t))} dt$. If $\E[
\Delta(\tilde p)  ]<+\infty$, then the inequalities
\[
\begin{split}
W_1(\Qxn,\QnN) & \leq
\frac{1}{\sqrt{N-n}} \,\,\E [ \Delta(\tp) | \xi(n) ]+\frac{ 2n }{N} \\
&\leq  \frac{1}{\sqrt{N-n}} Y+\frac{ 2n }{N},  \\
\end{split}
\]
 holds true
for all $n < N$, with $Y:=\sup_{n} \E[ \Delta(\tp) | \xi(n) ] <+\infty$, for $P$--almost every $\xi(n)$.
\end{proposition}

{\it Proof.} As already recalled,  $\sup_{(p,q) \in \P^2} \beta(p,q) \leq 2$ and $\beta$
satisfies $(\ref{qlin})$. 
Using the dual representation of $w_1$-- which is the analogue of
(\ref{dualrep}) with $\RE^d$ in the place of $\P$ and $\|\cdot\|$ in
the place of $\lm$-- it is easy to see that
\begin{equation}\label{BLvsW}
\beta(p,q) \leq w_1(p,q).
\end{equation}
Moreover,  recall that, when $X=\RE$,
\begin{equation}\label{w1}
w_1(p,q)=\int_\RE| F_p(x)-F_q(x)| dx.
\end{equation}
 See,
for instance,  \cite{Rachev1991}.
For any $p$ in $\P_1$, for  the sake of simplicity, set $\zi=z_i$ and observe that combination of  
(\ref{BLvsW}) and (\ref{w1}) gives
\[
\begin{split}
\E \left [ \lambda\Big(p, \frac{1}{N-n}\sum_{i=1}^{N-n} \delta_{z_i}
  \Big)  \right ] & \leq
\E \left [  \int_\RE|F_p(t)- \frac{1}{N-n} \sum_{i=1}^{N-n} \J(z_i \leq t)| dt \right ] \\
& = \frac{1}{N-n}  \E \left [ \int_\RE|(N-n)F_p(t)- \sum_{i=1}^{N-n}
\J(z_i \leq t)| dt \right ] \\ & \qquad (\text{by
 Fubini theorem}) \\
&= \frac{1}{N-n} \int_\RE
 \E\left [ | (N-n)F_p(t)-
\sum_{i=1}^{N-n} \J(z_i \leq t) | \right ] dt. \\&
\end{split}
\]
Now, note that $\sum_{i=1}^{N-n} \J(z_i \leq t)$ are binomial random
variables of parameters
 $((N-n), F_p(t))$. Hence, since
\begin{equation}\label{l22}
\int_{\RE} \sqrt{F_p(t)(1-F_p(t))}\, dt < +\infty
\end{equation}
holds true $P$-almost surely, from the Cauchy-Schwartz inequality one gets
\[
\E \left [ \lambda\Big(p, \nu_{N-n}^{(p)}
  \Big)  \right ] \leq \frac{1}{\sqrt{N-n}}
\int_{\RE} \sqrt{F_p(t)(1-F_p(t))} dt.
\]
Combination of this fact with Lemma \ref{L0} and the obvious
identity $\int_\P \Delta(p) \Qxn(dp)= \E [ \Delta(\tp) | \xi(n) ]$
gives the first part of the thesis. To conclude the proof,  apply
Doob's martingale convergence theorem (see, e.g., Theorem 10.5.1 in
\cite{Dudley2002}) to $ \E [ \Delta(\tp) | \xi(n) ] $ in order to
prove that $ \sup_n \E [ \Delta(\tp) | \xi(n) ] <+\infty$ almost
surely. $\quad \fine$

A first simple consequence of the previous proposition is 
embodied in 
\begin{corollary}
Let  $X=[-M,M]$ for some $0<M<+\infty$. Then,
\[
W_1(\Qxn,\QnN) \leq \frac{2M}{\sqrt{N-n}}+\frac{ 2n }{N}
\]
holds true for all $n < N$ for $P$--almost every $\xi(n)$.
\end{corollary}
It is worth  recalling that $\Delta(p)<+\infty$ implies finite
second moment for $p$ but not conversely (this condition defines the
Banach space $L_{2,1}$, cf. \cite{Ledoux1991}, p.10).
It is easy to show that if $p$ has  finite moment of order ${2+\delta}$, for some
positive $\delta$, then
\begin{equation}\label{1}
\Delta(p) \leq \left [1+ C_\delta \left(\int_\RE
|x|^{2+\delta}p(dx)\right)^{1/2} \right ]
\end{equation}
holds true with $C_\delta:=\sqrt{2(1+\delta)/\delta}$.
As a consequence of these statements we have the following

\begin{corollary} If
$\E[ \Delta(\tilde p)  ] <+\infty$, then
\[
\E [W_1(\Qxn,\QnN ) ] \leq \frac{\E[ \Delta(\tilde p)
]}{\sqrt{N-n}}+\frac{2n}{N}
\]
and
\[
P \left \{ W_1(\QnN,\Qxn) >\eps  \right \}\leq \frac{1}{\eps} \left
[\frac{\E[ \Delta(\tilde p)  ]}{\sqrt{N-n}}+ \frac{2n}{N}
   \right]
\]
hold true for all $n < N$. Moreover, if $\E|\xi_1|^{2+\delta}< +\infty$
for some positive $\delta$, then
\[
\E[ \Delta(\tilde p)  ]\leq \left[1+ \left
(\frac{2(1+\delta)}{\delta} \E|\xi_1|^{2+\delta}
\right )^{1/2} \right].
\]
\end{corollary}

{\it Proof.} By Proposition \ref{P2.2}, whenever $\E[ \Delta(\tilde
p)  ] < +\infty$, one can write
\[
\begin{split}
\E[  W_1(\Qxn,\QnN) ] &\leq \frac{1}{\sqrt{N-n}} \E \left[ \E
\big[\Delta(\tilde p)|\xi(n) \big]  \right ]+\frac{2n}{N}
\\
&= \frac{\E[ \Delta(\tilde p)  ]}{\sqrt{N-n}}+\frac{2n}{N} .
   \\
\end{split}
\]
Now, let  $\bar p(\cdot)=\E(\tilde p(\cdot))$. Then, (\ref{1}) together with
Fubini theorem and Jensen inequality  yield
\[
\E[ \Delta(\tilde p)  ] \leq    \left[1+C_\delta \left ( \int_{\RE}
|x|^{2+\delta}\bar p(dx) \right )^{1/2} \right].
\]
Combining these facts with Markov inequality completes the proof.
$\quad \fine$

\subsection{The case $X=\RE^d$} 
Let $X=\RE^d$
and $\lambda=\beta$. For any $p$ in $\P$ and $k$ in $\N$ consider
\[
\Psi_k(p):=\left (\sup_{\eps \in (0,1]} \eps^k
N(\eps,\eps^{k/(k-2)},p) \right )^{1/2}
\]
where  $N(\eps,\eta,p)$ is the minimal number of sets of diameter
$\leq 2\eps$ which cover $\RE^d$ except for a set $A$ with $p(A)
\leq \eta$. Proposition 3.1 in  \cite{Dudley1968} (see also Theorem
7 in \cite{Kalashnikov1990})  gives
\[
\E \left [ \beta\big(\,p\,, \,\nu_{N-n}^{(p)}\, \big )\right ] \leq
(N-n)^{-1/k} [\frac{4}{3}+ 4\cdot 3^{2k}\Psi_k(p) ].
\]
Using the last inequality and arguing as in the proof of Proposition
\ref{P2.2} we obtain the following
\begin{proposition}
If $\E[\Psi_k(\tilde p) ]<+\infty$ for some positive $k$, then the
inequality
\[
\begin{split}
W_1(\Qxn,\QnN) & \leq
\frac{1}{(N-n)^{1/k}} \left(\frac{4}{3}+4\cdot3^{2k}\E [\Psi_k(\tilde p)  | \xi(n) ]\right)+\frac{ 2n }{N} \\
&\leq \frac{1}{(N-n)^{1/k}} \left(\frac{4}{3}+4\cdot3^{2k}Y\right)+\frac{ 2n }{N},   \\
\end{split}
\]
 holds
true for all $n < N$, 
with $Y:=\sup_{n} \E[\Psi_k(\tilde p)  | \xi(n) ] <+\infty$,  for $P$--almost every $\xi(n)$.
\end{proposition}

\begin{remark}{\rm In the last proposition the fact that
$X=\RE^d$ does not play any special role. Everything remains true if $X$ is a
Polish space.}
\end{remark}

Condition $\E[\Psi_k(\tilde p) ]<+\infty$ is almost impossible to
check. In what follows we will assume a more tractable hypothesis.
If $\int_{\RE^d} \|x\|^{\gamma} p(dx)<+\infty$ where
$\gamma=\frac{kd}{(k-d)(k-2)}$, $d \geq 2$ and $k>d$, Proposition
3.4 in  \cite{Dudley1968} (see also Theorem 8 in
\cite{Kalashnikov1990})  yields
\begin{equation}\label{2}
\Psi_k(p)^2 \leq 2^d\left[1+ 2 \left(\int_{\RE^d}
\|x\|^{\gamma}p(dx)\right)^{1/\gamma} \right].
\end{equation}
Using this last inequality we can prove the following

\begin{proposition} Let $d \geq 2$, $k>d$ and set
$\gamma:=\frac{kd}{(k-d)(k-2)}$. Assume that  $ \E
\|\xi_1\|^{\gamma} $ is finite
and that $\gamma \geq 1$. If $Y_n:=2(
\E[\int_{\RE^d} |x|^{\gamma}\tp(dx)|\xi(n)] )^{1/\gamma}$, then, for
all $n < N$ and for $P$--almost every $\xi(n)$, one gets $Y:=\sup_n Y_n
<+\infty$ and
\begin{equation*}
W_1(\Qxn, \QnN) 
\leq \frac{1}{(N-n)^{1/k}} [ \frac{4}{3} + 4\cdot
3^{2k} \cdot 2^{d/2}   (1 +Y_n)^{1/2}] + \frac{2n}{N} 
\end{equation*}
Moreover,
\[
P \left \{ W_1(\Qxn,\QnN) >\eps  \right \}\leq \frac{1}{\eps} \left
[ \frac{K}{(N-n)^{1/k}} +\frac{2n}{N}
   \right]
\]
holds true for all $n < N$ with
\[
K=   \frac{4}{3} + 4\cdot 3^{2k} \cdot 2^{d/2} \big (1+2(
\E|\xi_1|^{\gamma})^{1/\gamma} \big)^{1/2} .
\]
\end{proposition}

{\it Proof.} Using (\ref{2}) and applying the Jensen inequality  two
times, we obtain
\[
\E[\Psi_k(\tilde p)| \xi(n)] \leq \{ 2^d + 2^{d+1} (\E[\int_{\RE^d}
\|x \|^{\gamma} \tilde p(dx)| \xi(n) ] )^{1/\gamma} \}^{1/2}.
\]
Combining Lemma \ref{1bound} with this last inequality, Doob's
martingale convergence theorem, Markov inequality and Jensen
inequality concludes the proof. $\quad \fine$

\subsection{Examples.}
The application  of the theorems of this section essentially require  conditions on the moments of $\xi_i$.
In the most common cases, 
the marginal distribution of each observation is available. Indeed, from a Bayesian point of 
view, the marginal distribution of each observation is usually treated as a prior guess of the mean of  the
unknown $\tilde p$. In the next three examples we review a few 
classical Bayesian nonparametric priors from this perspective.

\begin{example}[Normalized random measures with independent increments]{\rm
Probably the most celebrated example of nonparametric priors is the Dirichlet process, see, for example, 
\cite{Ferguson73,Ferguson74}.  A class of nonparametric priors which includes and generalizes 
the Dirichlet process  is the class of the so called {\it normalized random measures with independent increments}, introduced in  \cite{RLP} and studied, e.g., in \cite{Nieto2004,Lijoi2005b,Lijoi2005,lancillotto2005,Sangalli}. To define a normalized random measure with independent increments 
it is worth recalling that 
a random measure $\tm$ with independent increments  on  $\RE^d$ is  a random measure such that, for any measurable collection $\{A_1,\dots,A_k \}$ ($k \geq 1$) of pairwise disjoint measurable subsets of $\RE^d$, the random variable
$\tm(A_1),\dots, \tm(A_k)$ are stochastically independent.
Random measures with independent increments are completely characterized by
a measure $\nu$ on $\RE^d \times \RE^+$ via their Laplace functional.
More precisely, for every $A$ in $\CB(\RE^d)$ and every positive $\lm$
one has
\[
\E(e^{-\lm \tm(A)})=\exp \left \{-\int_{A \times \RE^+}(1-e^ {-\lm
v}) \nu(dx dv) \right \}.
\]
A systematic account of these random measures is given, for example, in \cite{Kingman67}.
Following \cite{RLP}, if \( \int_{\RE^d \times \RE^+} (1-e^ {-\lm v})
\nu(dx dv) < +\infty \) for every positive $\lm$ and \( \nu(\RE^d
\times \RE^+) = +\infty \), then one  defines a {\it normalized random
measure with independent increments}  putting  \(
\tp(\cdot):=\tm(\cdot)/\tm(\RE^d)\). In point of fact, under the
previous assumptions,   $P\{ \tm(\RE^d) = 0\}=0$; see \cite{RLP}.
The classical
example is the Dirichlet process, obtained with $\nu(dx dv)=\a(dx) \rho(dv)=\a(dx) v^{-1} e^{-v}
dv$, $\a$ being a finite measure on $\RE^d$.
Consider now a sequence  $(\xi_i)_{i \geq 1}$ of exchangeable random
variables driven by
$\tp$. 
When $\nu(dx dv)=\a(dx) \rho(dv) $, then $P\{ \xi_i \in  A\}=\a(A)/\a(\RE^d)$ for every
$i \geq 1$.
More generally, 
\[
P\{ \xi_i \in  A\}=\int_{\RE^{+}}
\phi(\lm)\int_{A
\times \RE^+}e^{-\lm u} u \nu(dx du) d \lm,
\]
where
\[
\phi(\lm):=\exp\{-\int_{\RE^k \times\RE^+}(1-e^{-\lm v})\nu(dy dv) \}
\]
see, e.g., Corollary 5.1 in \cite{Sangalli}. Hence, 
$\E \|\xi\|^m <+\infty $ if and only if 
\[
 \int_{\RE^{+}} \phi(\lm) \int_{\RE^k \times
\RE^+}e^{-\lm u} \|x\|^m u \nu(dx du) d\lm < +\infty.
\]
}
\end{example}

\begin{example}[Species sampling sequences and stick-breaking priors]{\rm
An exchangeable sequence of random variables $(\xi_n)_n$ is called a species sampling sequence 
(see \cite{Pitman96}) if, for each $n\geq 1$,  
\[
P\{\xi_{n+1} \in A| \xi(n) \}=l_{0,n} \a(A) + \sum_{j=1}^{k(n)} l_{j,n} \delta_{\xi^*_j}(A) \qquad (A \in \CX)
\]
and 
\[
P\{ \xi_1 \in A\}=\a(A)
\]
with the proviso that $\xi^*_1,\dots \xi^*_{k(n)}$ are the $k(n)$ distinct values of $\xi_1,\dots,\xi_n$ in the same 
 order as they appear,
$l_{j,n}$ ($j=0,\dots,k(n)$) are non-negative measurable functions of $(\xi_1,\dots,\xi_n)$, 
and $\a$ is some non--atomic probability measure on $(X,\CX)$. See, among others, \cite{Pitman2000,Pitman2005,Pitman95,pitman2003}.
Of course, in this case, it a simple task  to check conditions on the marginal distribution of each observation,  
since it coincides with $\a$.
A particular kind of random probability laws connected with the species sampling sequences are the so-called
stick-breaking priors. Such priors are almost surely discrete random probability
measures that can be represented as
\[
\tilde p(\cdot)= \sum_{k=1}^{N} p_k \delta_{Z_k}(\cdot)
\]
where $(p_k)_{k \geq 1}$ and $(Z_k)_{k \geq 1}$ 
are independent, $0 \leq p_k \leq 1$ and $\sum_{k=1}^{N} p_k=1$ almost surely,
and $(Z_k)_{k \geq 1}$ are independent and identically distributed random variable taking values in $X$
with common probability $\a$. Stick-breaking priors can be constructed using either a finite or infinite numbers of 
terms, $1 \leq N \leq +\infty$. Usually,
\[
p_1=V_1 \qquad p_k=(1-V_1)(1-V_2)\dots (1-V_{k-1})V_k \qquad k \geq 2
\]
where $V_k$ are independent $Beta(a_k,b_k)$ random variables for $a_k >0,b_k>0$. See \cite{Ishwaran1,Ishwaran2}.
It is clear that in this case
\[
P\{ \xi_i \in A\}= \a(A).
\]}
\end{example}

\newcommand\genit[1]{\operatorname{\rm{ge}}(#1)}
\begin{example}[Pólya tree]{\rm 
Let $X=\RE$ and let  $E_j$  be the set of all sequences of $0$s and $1$s of length $j$. Moreover, set
$E^*=\cup_jE_j$. For each $n$, let $\CT_n=\{B_{\bar \eps}: \bar \eps \in E_n \}$ be a partition of $\RE$ such that 
for all $\bar \eps $ in $E^*$, $B_{\bar \eps 0}$, $B_{\bar \eps 1}$ is a partition of $B_{\bar \eps}$.   
Finally let $\aleph=\{\a_{\bar \eps}: \bar \eps \in E^*\}$ be a set of nonnegative real numbers. 
A random probability $\tilde p$ on $\RE$ is said to be a P\'olya tree with respect to the partition $\CT=\{ \CT_n\}_n$
with parameter $\aleph$ if
\begin{itemize}
\item $\{\tilde p\,(\,B_{\bar \eps 0}|B_{\bar \eps 0}): \bar \eps \in E^* \}$ are a set of independent random variables
\item  for all $\bar \eps$ in $E^*$  $\tilde p\,\,(B_{\bar \eps 0}|B_{\bar \eps 0})$ is 
$Beta(\a_{\bar \eps 0},\a_{\bar \eps 1})$.
\end{itemize}
See  \cite{MW90,Lavine92,Lavine94}.
Under suitable condition on $\aleph$, such a random probability does exist. See Theorem 3.3.2 in \cite{Rama}.
Moreover, if $(\xi_n)_{n \geq 1}$ is an exchangeable sequence with driving measure $\tilde p$, 
for any $B_{\bar \eps}$ with $\bar \eps=\eps_1 \eps_2 \dots \eps_k$, 
\[
P\{ \xi_n \in B_{\bar \eps}\}= \prod_{i=1}^k \frac{\a_{\eps_1 \eps_2 \dots \eps_i }}{\a_{\eps_1 \eps_2 \dots \eps_i 0}
+\a_{\eps_1 \eps_2 \dots \eps_i 1} }.
\]
See, e.g., Theorem 3.3.3 in \cite{Rama}. In this case it is a difficult task to give explicit conditions for the existence 
of the moments of $\xi_i$. Nevertheless, Lavine suggests that, if the partitions has the form
\(
F^{-1}(\sum \eps_i/2^i,\sum \eps_i/2^i+1/2^i),
\)
 $F$ being a continuous distribution function, and 
\[
\frac{\a_{\eps_1 \eps_2 \dots \eps_i }}{\a_{\eps_1 \eps_2 \dots \eps_i 0}
+\a_{\eps_1 \eps_2 \dots \eps_i 1} }=\frac{1}{2}
\]
then $P\{\xi_n \leq x\}=F(x)$. 
}
\end{example}

\section*{Acknowledgments}
I am grateful to Eugenio Regazzini for providing much of the inspiration behind this paper.
Moreover I want also to thank Luca Monno, who is a virtual coauthor of this paper, 
and Laura Sangalli for helpful comments.
This work was partially supported  by the IMATI (CNR - Pavia, Italy).



\bibliographystyle{plain}

\end{document}